\documentclass[fleqn,usenatbib]{mnras}
\usepackage{newtxtext,newtxmath}
\usepackage[T1]{fontenc}
\DeclareRobustCommand{\VAN}[3]{#2}
\let\VANthebibliography\thebibliography
\def\thebibliography{\DeclareRobustCommand{\VAN}[3]{##3}\VANthebibliography}
\usepackage[flushleft]{threeparttable}
\usepackage{graphicx}	
\graphicspath{ {./figures/} }
\usepackage{amsmath}	
\usepackage{tabularx}
\usepackage{makecell, multirow}

\title[Model-independent Redshift Estimation of BL Lac Objects]{Model-independent Redshift Estimation of BL Lac Objects through VHE Observations}

\author[Malik Zahoor et al.]{
Malik Zahoor$^{1}$\thanks{E-mail: malikzahoor313@gmail.com},
Sunder Sahayanathan$^{2,3}$\thanks{E-mail: sunder@barc.gov.in},
Shah Zahir$^{4,5}$,
Naseer Iqbal$^{1}$,
Aaqib Manzoor$^{1}$ \&
Nilay Bhatt$^{2}$
\\
$^{1}$Department of Physics, University of Kashmir, Srinagar 190006, India.\\
$^{2}$Astrophysical Sciences Division, Bhabha Atomic Research Center, Mumbai 400085, India.\\
$^{3}$Homi Bhabha National Institute, Mumbai 400094, India.\\
$^{4}$Inter-University Center for Astronomy and Astrophysics, Post Bag 4, Ganeshkhind, Pune 411007, India.\\
$^{5}$Department of Physics, Central University of Kashmir, Ganderbal 191201, India. 
}

\date{Accepted XXX. Received YYY; in original form ZZZ}

\pubyear{2021}

\begin{document}
\label{firstpage}
\pagerange{\pageref{firstpage}--\pageref{lastpage}}
\maketitle

\begin{abstract}
The very high energy (VHE) gamma-ray spectral indices of blazars show strong correlation with the source redshift. Absence of any such correlation in low energy gamma rays and X-rays indicate the presence of Extragalactic Background Light (EBL) induced absorption of VHE gamma rays. By employing a linear regression analysis, this observational feature of blazars is used to constrain the redshift of BL Lac objects which was unknown/uncertain earlier. Additionally, we also compare the observed VHE spectral index-redshift correlation with the ones predicted from commonly adopted EBL models. Our study highlights the deviation of the EBL model based predictions from the observation especially at high redshifts.  
\end{abstract}



\section{Introduction}
\label{sec1}
The astral sky in the VHE regime ($E > 100$ GeV) is dominated by blazars, a subclass of radio loud active galactic nuclei (AGN) with a relativistic jet pointing at a small angle with respect to line of sight \citep{1995PASP..107..803U}. Due to this arrangement, the measured flux over the whole electromagnetic spectrum gets Doppler boosted with apparent variability as short as few minutes \citep{Gaidos,2007ApJ...669..862A,Aharonian_2007,Arlen_2012}. The spectral energy distribution (SED) of blazars is characterised by two broad emission components: the low energy component peaking in the optical-to-X-ray band and the high energy peak located in the gamma-ray bands \citep{1995PASP..107..803U}. The first component is believed to be the synchrotron emission from a relativistic distribution of electrons in the jet, while the emission process responsible for the second component is still under debate. Under leptonic scenario, the high energy component is explained by inverse Compton (IC) scattering of either the synchrotron (synchrotron self Compton - SSC:  \citep{1965ARA&A...3..297G,1998ApJ...509..608T}) or external (external Compton - EC: \citep{1993ApJ...416..458D}) photons by electrons and positrons present in the jet. On the contrary, various hadronic models explain this component through proton-synchrotron emission and/or nuclear cascades \citep{2003APh....18..593M, 2015A&A...573A...7W}. 

Blazars are further classified as BL Lac objects and flat spectrum radio quasars (FSRQs), with the later exhibiting significant emission and/or absorption lines while such features are absent/weak in the former. Based on the energy at which the synchrotron spectral component peaks  $\nu_{sy}$, blazars are sub-divided into low-frequency peaked blazars (LBL, $\nu_{sy}<10^{14}$ Hz), intermediate-frequency peaked blazars (IBL, $10^{14}\leq\nu_{sy}<10^{15}$ Hz), high-frequency peaked blazars (HBL, with $10^{15}\leq\nu_{sy}\leq10^{17}$ Hz) and extreme high-frequency peaked blazars (EHBL, $\nu_{sy}>10^{17}$ Hz) \citep{10.1093/mnras/stz812}. While FSRQs are all LBLs, BL Lac objects fall under different categories. Most of the blazars detected by ground based Imaging Atmospheric Cherenkov telescope (\emph{IACT}) facilities in very high energy gamma-rays (VHE; $E \geq 100$ GeV) are BL Lac objects belonging to the HBL/EHBL class. EHBLs are relatively less luminous compared to other HBLs and their synchrotron spectral component peaks at energy $> 10$ keV.

Unlike other wavebands, VHE gamma ray emission form blazars undergo significant en route attenuation due to pair production losses with the Extragalactic Background Light (EBL) \citep{1967PhRv..155.1404G}. EBL is the diffuse radiation with main contributors as the light from galaxies and the reprocessed dust emission of the universe. Accordingly, the EBL spectrum peaks at UV and IR wavelengths and contain information that trace back to the structure formation epoch of our cosmic evolution \citep[see e.g.,][]{doi:10.1146/annurev.astro.39.1.249, DWEK2013112}. Direct measurement of EBL is heavily hampered by the presence of strong zodiacal and galactic emission, and hence indirect estimates involving cosmological models are employed. Such estimates broadly fall under two categories namely backward evolution models and forward evolution models. 
In the former approach, one begins with the existing galaxy population and extrapolates back in time using cosmological 
models to estimate the EBL \citep{Franceschini_2017, refId8, Stecker_2006, 1999ApJ...522..604P}. The later approach begins
with cosmological initial conditions and evolves with time. Here, the model parameters are adjusted to reproduce the 
observed property of the current universe \citep{Inoue_2013, 10.1111/j.1365-2966.2012.20841.x, Primack_2005}. Besides these 
approaches, some cosmological models use the SED of galaxy stellar population and combine it with the cosmic star formation 
history to calculate the EBL intensity \citep{Khaire2015STARFH, Kneiske, Finke_2010}. The EBL intensity estimated under 
these models heavily depends on the adapted cosmological conditions and may vary substantially once the underlying assumptions are relaxed \citep{DWEK2013112}.

The attenuation introduced by the EBL cause the observed VHE spectrum of distant blazars to be significantly different from the 
source spectrum. In other words, the VHE spectra of blazars carry the signature of EBL and can be used as a probe/test the 
cosmological EBL models. However, the uncertainty regarding the EBL and the intrinsic VHE spectra of blazars makes the problem 
self consistent. Broadband spectral modelling of blazars using different emission processes can be used to predict the intrinsic VHE spectra and this in turn can provide constraints on various EBL models \citep{Mankuzhiyil_2010}. Nevertheless, the number of free parameters deciding the broadband spectra of blazars are significantly large to be constrained and the ambiguity regarding the nature of the radiating particle is a major hindrance for a realistic estimate of the intrinsic VHE spectrum. Additionally, the 
detection of the gravitationally lensed blazar S3\,0218+35 at relatively high redshift of $z = 0.944$ \citep{refId0} poses serious 
challenges to the existing EBL models and the nature of the intrinsic VHE spectrum of blazars.

The problem can be recast to estimate the distance of BL Lac objects which lack prominent line 
emissions \citep[see e.g.,][]{1999ApJ...525..127L}. Assuming certain EBL model to be valid over large distances and with an 
acceptable estimate of intrinsic VHE spectrum, an innovative way to estimate source distance can be employed by reproducing the 
observed VHE spectrum \citep[see e.g.,][]{2010ApJ...708L.100A, Archer_2018}. However, the uncertainty regarding the EBL intensity 
compounded with the model dependent estimates of the intrinsic VHE spectrum hampers the reliability of this distance estimation, 
particularly for distant sources. In this work, we propose a novel method to calculate the redshift of the BL Lac objects by exploiting the positive correlation between the observed VHE spectral indices and the source redshifts of blazars for which better estimates of their distance is available. Particularly, this redshift estimate do not depend on the assumptions of EBL model and the 
intrinsic VHE spectra of blazars. We further extend the study to highlight the discrepancy of four commonly used EBL models 
at high redshifts.

The paper is organised as follows: In the following section, we perform a detailed correlation study between the spectral index 
and redshift of blazars followed by the linear regression analysis. In section \S \ref{sec3}, we use the regression relations to 
estimate the redshift of 6 BL Lac objects. The comparison between the predicted VHE spectral index-redshift relation by the EBL models with the observation is presented in section \S \ref{sec4}. Finally, the results of the present work are discussed and 
summarised in section \S \ref{sec5}. Throughout this work we adapt a cosmology with $\Omega_M$ = 0.3, $\Omega_\Lambda$ = 0.7, and 
$H_0 = 71 km s^{-1} Mpc^{-1}$ .

\section{EBL signature in VHE spectrum of blazars}
\label{sec2}
The presence of EBL induced attenuation in VHE spectra of distant blazars can be readily understood from the positive correlation between the observed VHE indices with redshift of distant blazars. This was first systematically shown by \citet{Sinha_2014} considering 29 blazars belonging to HBL and EHBL classes. Absence of significant correlation between the spectral indices in other waveband with the source redshift further asserts this inference. We updated the results of \citet{Sinha_2014} by including the recent blazars detected at VHE.
This includes additional 3 HBLs and 5 EHBLs. We have also excluded IC\;310 from the list of EHBL since there are no substantial 
evidence to confirm its class. Additionally, we have also included 7 FSRQs in the present study. The sources selected for the present study are listed in {\it TevCat}\footnote{\url{http://tevcat.uchicago.edu/}} which include all the blazars detected by \emph{HESS}, \emph{MAGIC}, \emph{VERITAS} and \emph{WHIPPLE}. For the sources with more than one observation, we selected the VHE spectral index ($\Gamma$) corresponding to its lowest flux state. This bias will have only a negligible effect on our study results since the variation in spectral index for an individual source is much less compared to the spectral steepening introduced by EBL (section \S \ref{sec5}). Nevertheless, since the blazar spectrum is known to be highly variable and to quantify this variation we study the standard deviation in the VHE spectral index of two well observed BL Lac objects Mkn\,421 and Mkn\,501. For Mkn\,421, this was found to be $\approx$ 0.21 evaluated independently from the 14 observations by \emph{MAGIC} and 8 by \emph{VERITAS} \citep{refId9, Balokovi__2016}, while in case of Mkn\,501 the standard deviation in VHE spectral index was $\approx$ 0.14 estimated from 15 observations by \emph{MAGIC} and 3 by \emph{VERITAS} \citep{501_m1,501_m2}. This accounts to $\approx 9\%$ VHE spectral index variation for Mkn\,421 and $\approx 7 \%$ for Mkn\,501. We assume similar percentage of variation in the VHE spectral index to be present for all BL Lacs and hence, $9\%$ additional error is applied along with the observational errors. The number of VHE observations for any FSRQ is too less to repeat a similar treatment and hence we consider only the observational errors for the VHE 
spectral index. Besides this, we also consider for the present study the average VHE spectral index ($\Gamma_{av}$) 
estimated from all available VHE observations for HBLs and EHBLs.

The en route absorption of VHE photons by the EBL through pair production process results in the steepening of the observed VHE spectrum of blazars \citep[see e.g.,][]{2000APh....12..217V, Mankuzhiyil_2010}. This effect will be more pronounced for distant sources than the nearer ones. Hence, one would expect a correlation between the observed VHE spectral indices with the source redshift, provided the sources are assumed to have similar intrinsic VHE spectral indices. To investigate this, we perform Pearson and Spearman rank correlation studies between observed VHE spectral index\footnote{We define the spectral index, $\Gamma$, such that dN/dE $\propto$ E$^{-\Gamma}$ (photon cm$^{-2}$ s$^{-1}$ TeV$^{-1}$).} and the source redshift for the blazars listed in Table~\ref{tab1}. Since the intrinsic VHE index of HBLs can be harder than FSRQs \citep{Ackermann_2011}, we perform the correlation study on individual classes rather than the entire sample. The list of HBLs and FSRQs are given in the top and bottom panel of Table~\ref{tab1}. In the middle panel, we provide the list of EHBLs for which the intrinsic VHE spectrum obtained considering various EBL models is extremely hard with an index $< 2$ \citep[see e.g.,][]{2014MNRAS.438.3255T, 2014ApJ...787..155T}.

\subsection{Spectral index -- Redshift Correlation}
\label{sec2.1}
\begin{figure}
		\centering
		\includegraphics[scale=0.3, angle=270]{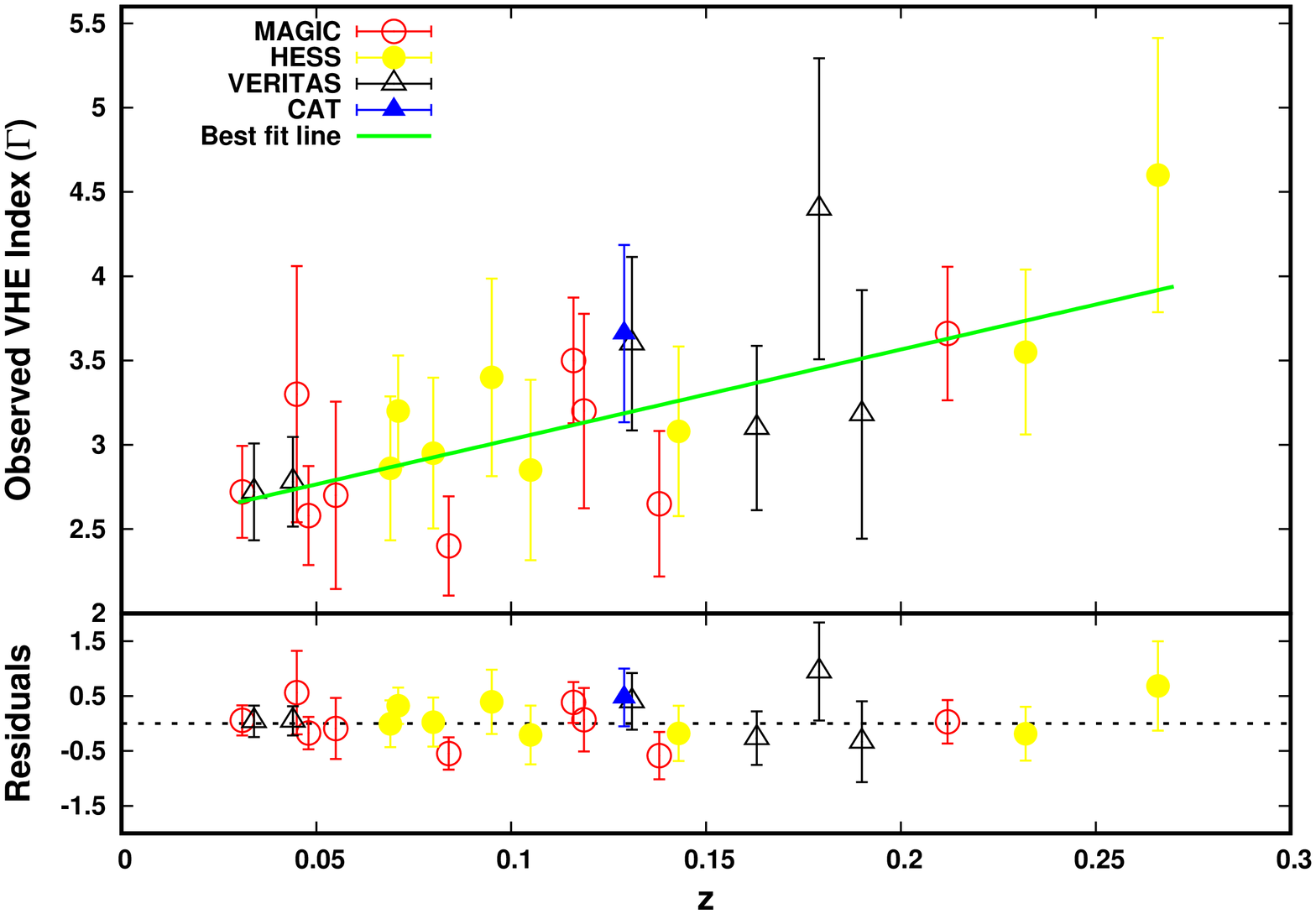}
		\caption{\textbf{\emph{Top}}: Scatter plot between observed VHE spectral index of the selected HBLs with redshift. Different color/symbol depict observations from different VHE telescopes as mentioned in labels. The solid line (green) is the best-fit straight line to the HBLs with $\chi^2/{\rm d.o.f} = 12.97/22$, \textbf{\emph{Bottom}}: Residual plot.}
		\label{Fig1}
\end{figure}

\begin{figure}
		\centering
		\includegraphics[scale=0.3, angle=270]{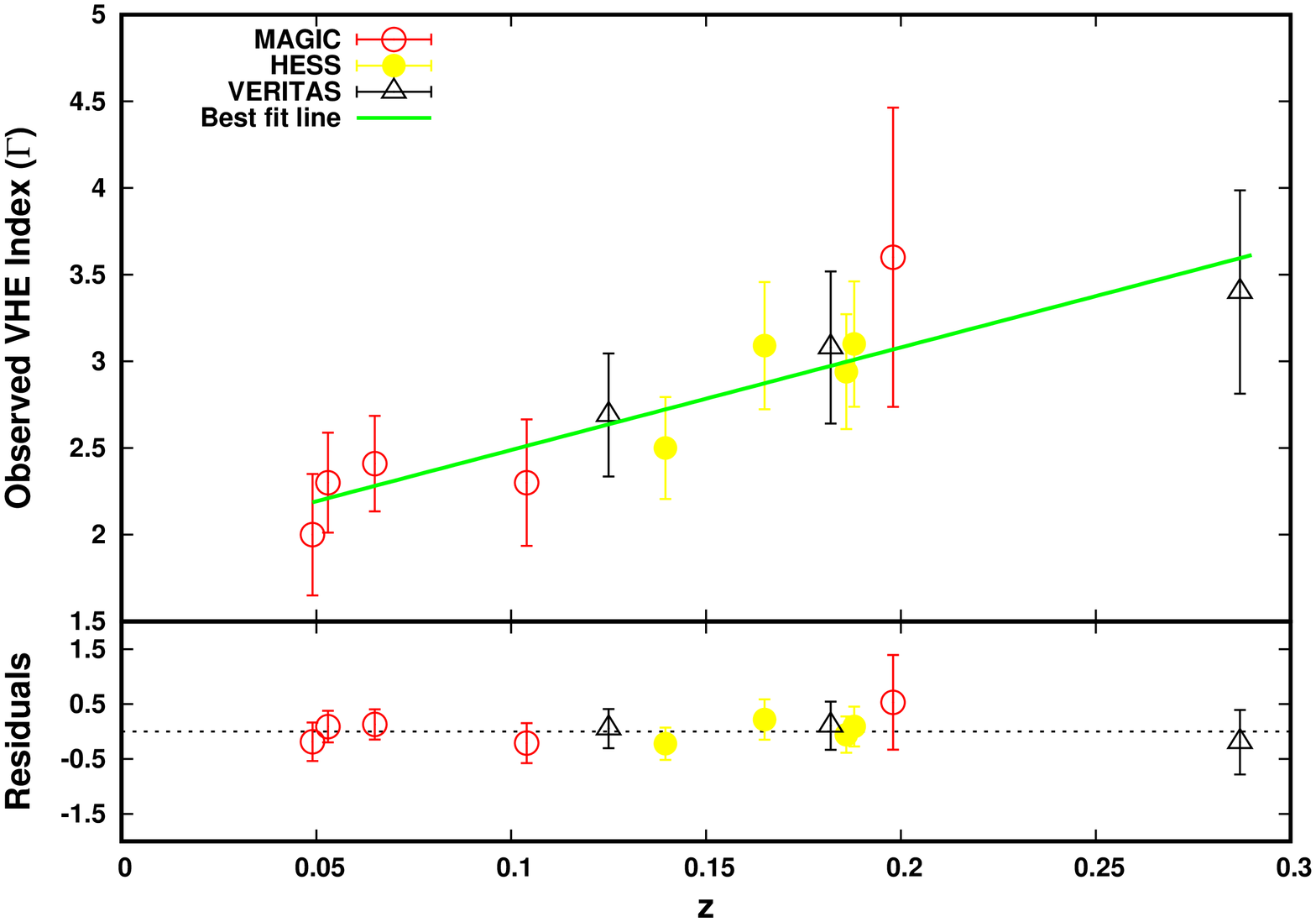}
		\caption{\textbf{\emph{Top}}: Scatter plot between observed VHE spectral index of the selected EHBLs with redshift. Different color/symbol depict observations from different VHE telescopes as mentioned in labels. The solid line (green) is the best-fit straight line to the EHBLs with $\chi^2/{\rm d.o.f} = 2.52/10$, \textbf{\emph{Bottom}}: Residual plot.}
		\label{Fig2}
\end{figure}

\begin{figure}
		\centering
		\includegraphics[scale=0.3, angle=270]{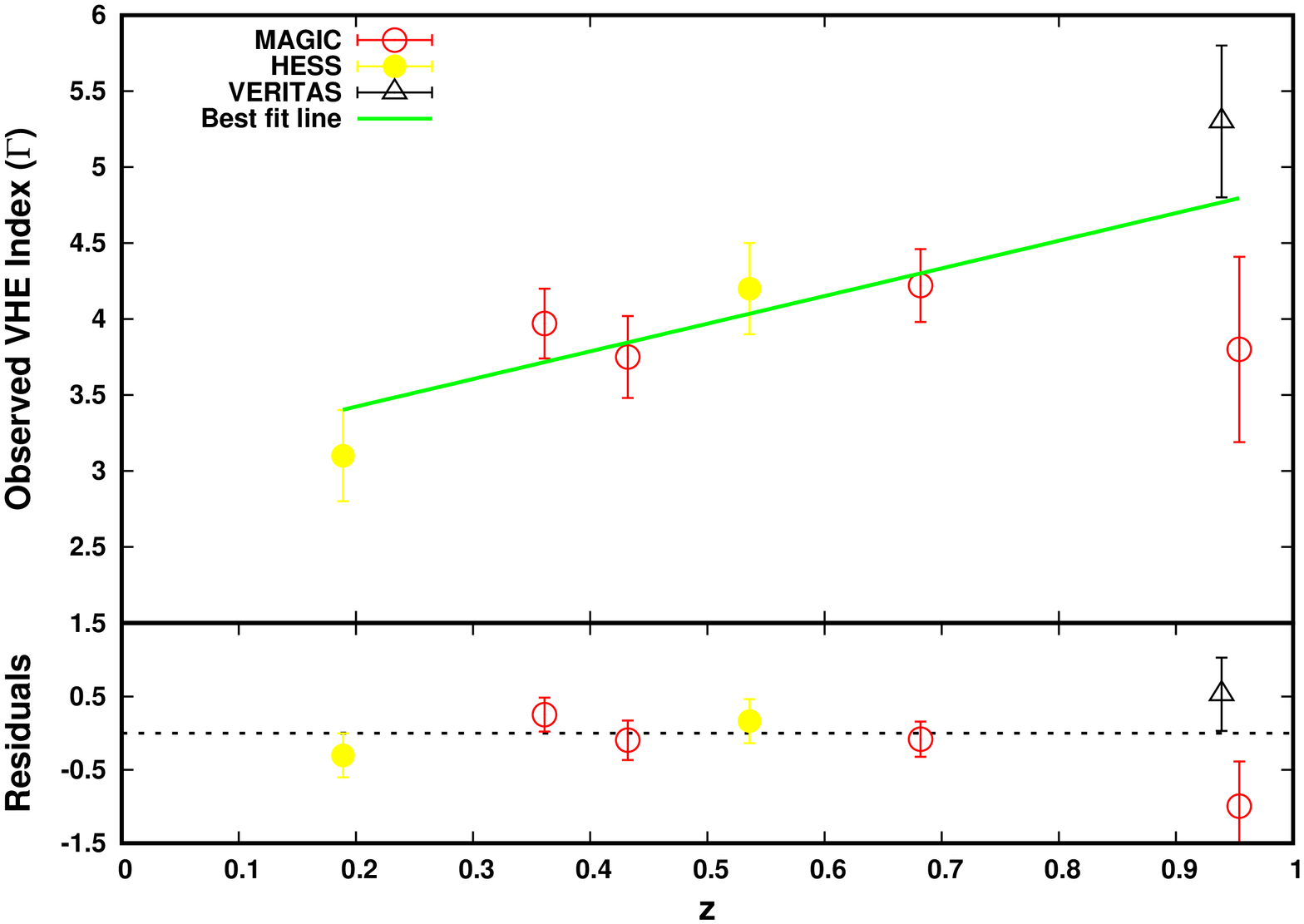}
		\caption{\textbf{\emph{Top}}: Scatter plot between observed VHE spectral index of the FSRQs with redshift. Different color/symbol depict observations from different VHE telescopes as mentioned in labels. The solid line (green) is the best-fit straight line to the FSRQs with $\chi^2/{\rm d.o.f} = 6.57/5$, \textbf{\emph{Bottom}}: Residual plot.}
		\label{Fig3}
\end{figure}

To examine the dependence of the observed VHE spectral index ($\Gamma$) on the redshift ($z$), we perform a correlation study between these quantities for the ensemble of blazars listed in Table~\ref{tab1}. The Spearman rank correlation analysis for the case of 
HBLs resulted in rank correlation coefficient, $r_s = 0.63$ with the null hypothesis probability of $P_{rs}< 0.05$. Similarly, the Pearson's correlation analysis resulted in the linear correlation coefficient $\rho = 0.72$ with the null hypothesis probability $P_{\rho} < 0.05$. In Figure~\ref{Fig1}, we show the scatter plot between $\Gamma$ and $z$ for the selected HBLs. For EHBLs the dependence between $\Gamma$ and $z$ is more pronounced with $r_s = 0.94$ and $P_{rs}< 0.05$, while $\rho = 0.89$ and $P_{\rho} < 0.05$. The scatter plot between $\Gamma$ and $z$ for EHBLs is shown in Figure~\ref{Fig2}. In case of FSRQs, the correlation is moderate/inconclusive and the results are $r_s = 0.57$ with $P_{rs} = 0.18$ and $\rho = 0.69$ with $P_{\rho} = 0.087$. The correlation improves to $r_s = 0.94$ with $P_{rs} = 0.004$ and $\rho = 0.95$ with $P_{\rho} = 0.002$ when the high redshift blazar with spiral morphology, S3\,0218+35 ($z=0.954$) is omitted. However, with only 7 FSRQs detected at VHE energies this correlation should be treated only as indicative. The corresponding scatter plot for the FSRQs is given in Figure~\ref{Fig3}. When the correlation study is repeated with $\Gamma_{av}$, we found the the results are $r_s = 0.67$ with  $P_{rs}< 0.05$ and $\rho = 0.75$ with $P_{\rho} < 0.05$ for HBLs, and $r_s = 0.95$ with  $P_{rs}< 0.05$ and $\rho = 0.89$ with $P_{\rho} < 0.05$ for EHBLs.

\begin{figure}
		\centering
		\includegraphics[scale=0.3, angle=270]{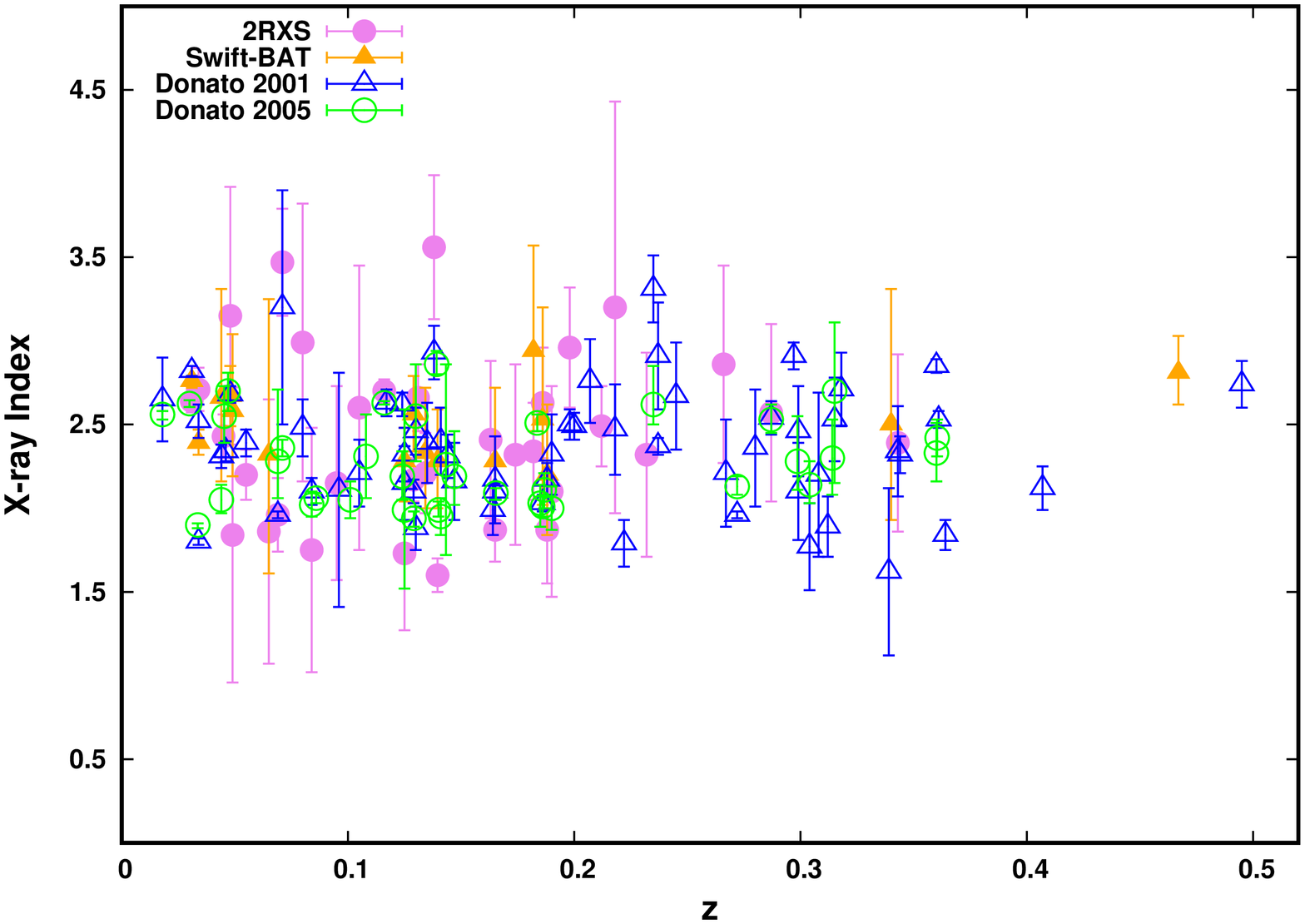}
		\caption{Scatter plot between observed X-ray spectral index of HBLs with redshift. The filled violet circles are from the \emph{ROSAT} catalog \citep{Boller_2016}, the yellow filled triangles from the \emph{Swift}-BAT catalog \citep{Oh_2018}), the blue triangles from archival X-ray catalog \citep{refId2}, and the green circles from the \emph{Beppo}-SAX catalog \citep{article}.}
		\label{Fig4}
\end{figure}

Our study illustrates that there exist a definite correlation between the observed VHE spectral index with the source redshift. This can be attributed to the EBL induced absorption provided the correlation is not associated with the cosmological evolution of blazars. The later presumption can be tested by performing a correlation study between spectral index at lower energies and the redshift. A positive correlation can validate the cosmological evolution or alternatively, falsify the signature of EBL induced absorption on the VHE spectra of distant blazars. We study the correlation between X-ray spectral indices with redshift for HBLs using: 105 months of the \emph{Swift}-BAT catalog consisting of 16 HBLs \citep{Oh_2018}, second \emph{ROSAT} all-sky survey (2RXS) source catalog containing 34 HBLs \citep{Boller_2016}, six years of the \emph{Beppo}-SAX catalog consisting of 38 HBLs \citep{article}, and an archival X-ray catalog from \emph{ASCA}, \emph{EXOSAT}, \emph{Beppo}-SAX, \emph{ROSAT} and \emph{EINSTEIN} consisting of 61 HBLs \citep{refId2} (Figure~\ref{Fig4}). The Spearman rank correlation study results are: $r_s = -0.15$ with $P_{rs} = 0.55$ for Swift-BAT, $r_s = 0.04$ with $P_{rs} = 0.78$ for 2RXS, $r_s = 0.02$ with $P_{rs} = 0.88$ for \emph{Beppo}-SAX and $r_s = -0.03$ with $P_{rs} = 0.79$ for the archival X-ray catalog. Absence of appreciable correlation between the X-ray spectral index and the redshift supports the presence of EBL signature in the VHE spectra of blazars. Additionally, we perform the the Spearman rank correlation between the low-energy gamma-ray (GeV) spectral index and redshift for the 44 HBLs listed in the fourth catalogue of \emph{Fermi}-LAT \citep{Abdollahi_2020}. We obtained $r_s = -0.15$ with $P_{rs} = 0.85$ suggesting a poor correlation and this further supports the steepening of the VHE spectral index with redshift to be an outcome of EBL induced absorption. Lack of correlation between the spectral index and luminosity in VHE further asserts the correlation between VHE spectral index and redshift is not associated with the Malmquist bais \citep{Sinha_2014}.

\begin{table*}
\centering
\begin{tabular}{|l|c|c|c|c|c|c|c|}
\hline 
Source Name & Type & z & $\Gamma$ & E$_ {VHE}$ & Total no. of VHE obs & $\Gamma_{av}$ & Ref \\ 
\hline 
Markarian\;421  	&HBL	   & 0.031	&	2.72 $\pm$ 0.12 & $>0.2$ & 22 & 2.74 $\pm$ 0.56 & \citet{refId9} \\
Markarian\;501  	&HBL	   & 0.034	&	2.72 $\pm$ 0.15 & $>0.1$ & 18 & 2.30 $\pm$ 0.51 & \citet{Acciari_2011} \\	
1ES\;2344+514	&HBL	   & 0.044	&	2.78 $\pm$ 0.09 & $0.39-8.3$ & 05 & 2.65 $\pm$ 0.27 & \citet{Acciari:2011jw}\\
PKS\;2155-304	&HBL	   & 0.116	&	3.5  $\pm$ 0.2 & $>0.4$ & 02 & 3.51 $\pm$ 0.20 & \citet{Aleksic:2012hd}\\
1ES\;1959+650	&HBL	   & 0.048	&	2.58 $\pm$ 0.18 & $0.18-2$ & 04 & 2.82 $\pm$ 0.31 & \citet{Tagliaferri_2008}\\
H\;1426+428	    &HBL      & 0.129	&	3.66 $\pm$ 0.41 & $>0.25$ &  03 & 3.25 $\pm$ 0.80 & \citet{refId11}\\
PKS\;2005-489	&HBL    & 0.071	&	3.20 $\pm$ 0.16 & $0.3-5$ & 02 & 3.60 $\pm$ 0.43 & \citet{2010}\\				
Markarian\;180   &HBL      & 0.045	&	3.3	 $\pm$ 0.7 & $>0.25$ &  01 & ... & \citet{article1}\\	
PKS\;0548-322	&HBL      & 0.069	&	2.86 $\pm$ 0.34 & $>0.25$ & 02 & 2.83 $\pm$ 0.45 & \citet{Aharonian_2010}\\
1ES\;1011+496	&HBL	&    0.212	&	3.66 $\pm$ 0.22 & $>0.15$ & 04 & 3.49 $\pm$ 0.74 & \citet{Aleksi__2016}\\
RGB\;J0152+017	&HBL	&    0.08	&	2.95 $\pm$ 0.36 & $>0.3$ & 01 & ... & \citet{Aharonian_2008}\\
1ES\;0806+524	&HBL	 &   0.138	&	2.65 $\pm$ 0.36 & $>0.3$ & 02 & 3.12 $\pm$ 1.06 & \citet{Aleksi__2015}\\
RBS\;0413		&HBL     &  0.19	&	3.18 $\pm$ 0.68 & $>0.25$ &  03 & 3.19 $\pm$ 0.97 & \citet{Aliu_2012i}\\			
1ES\;1440+122	&HBL	   & 0.163	&	3.1	 $\pm$ 0.4 & $0.2-1.3$ & 02 & 3.25 $\pm$ 0.80 & \citet{Archambault_2016}\\
RX\;J0648+1516	&HBL    &	0.179	&	4.4 $\pm$ 0.80 & $>0.2$ & 01 & ... & \citet{Aliu_2011}\\  
B3\;2247+381		&HBL     &  0.118	&	3.2	 $\pm$ 0.5 & $>0.2$ & 01 & ... & \citet{Aleksi__2012}\\
SHBL\;J001355.9-185406 &HBL &0.095	&	3.4	 $\pm$ 0.5 & $>0.31$ &  01 & ... & \citet{2013i}\\  	
1RXS\;J101015.9-311909 &HBL &0.143	&	3.08 $\pm$ 0.42 & $>0.2$ & 02 & 3.09 $\pm$ 0.65 & \citet{2012}\\
1ES\;1312-423	&HBL   	&0.105	&	2.85 $\pm$ 0.47 & $>0.28$ & 01 & ... & \citet{2013ii}\\
1ES\;1215+303	&HBL    	&0.131	&	3.6	 $\pm$ 0.4 & $>0.2$ & 03 & 3.38 $\pm$ 0.65 & \citet{Aliu_2013}\\
1ES\;1741+196	&HBL   	&0.084	&	2.4	 $\pm$ 0.2 & $0.08-3$ & 02 & 2.55 $\pm$ 0.72 & \citet{Ahnen_2017}\\   		    
1ES\;1727+502	&HBL    	&0.055	&	2.7	 $\pm$ 0.5 & $>0.15$ & 03 & 2.40 $\pm$ 0.58 & \citet{Aleksi__2014}\\ 	
PKS\;0301-243   	&HBL    	&0.266	&	4.6	 $\pm$ 0.7 & $>0.2$ & 01 & ... & \citet{2013iii}\\ 		   
1RXS\;J023832.6-311658 &HBL	&0.232	&	3.55 $\pm$ 0.371 & $>0.1$ & 01 & ... & \citet{2017ICRC...35..645G}\\
\hline
1ES\;1101-232  &EHBL &0.186   &   2.94 $\pm$ 0.20 & $>0.22$ & 01 & ... & \citet{Aharonian_2007i}\\
H\;2356-309    &EHBL &0.165   &   3.09 $\pm$ 0.24 & $0.2-1.3$ & 02 & 3.07 $\pm$ 0.28 & \citet{Aharonian_2006}\\
1ES\;1218+304  &EHBL &0.182   &   3.08 $\pm$ 0.34	& $0.16-1.8$ &  02 & 3.04 $\pm$ 0.52 & \citet{Fortin_2008}\\
1ES\;0229+200  &EHBL &0.1396  &   2.50 $\pm$ 0.19	& $0.5-15$ & 02 & 2.46 $\pm$ 0.21 & \citet{Aharonian_2007ii}\\
1ES\;0347-121  &EHBL &0.188   &   3.10 $\pm$ 0.23	& $0.25-3$ & 01 & ... & \citet{Aharonian_2007iii}\\
RGB\;J0710+591 &EHBL &0.125   &   2.69 $\pm$ 0.26	& $$>0.3$$ & 01 & ... & \citet{Acciari_2010}\\
1ES\;0414+009  &EHBL &0.287   &   3.4 $\pm$  0.5	& $0.23-0.85$ & 03 & 3.42 $\pm$ 0.62 & \citet{Aliu_2012ii}\\
RBS\;0723      &EHBL &0.198   &   3.60 $\pm$ 0.80	& $>0.1$ &  01 & ... & \citet{MAGIC}\\
1ES\;2037+521  &EHBL &0.053   &   2.30 $\pm$ 0.2	& $>0.1$ & 01 & ... & \citet{MAGIC}\\
PGC\;2402248   &EHBL &0.065   &   2.41 $\pm$ 0.17	& $>0.22$ & 01 & ... & \citet{2019}\\
TXS\;0210+515  &EHBL &0.049   &   2.0 $\pm$ 0.30	& $>0.1$ & 02 & 1.95 $\pm$ 0.41 & \citet{MAGIC}\\ 
RGB\;J2042+244 &EHBL &0.104   &   2.3 $\pm$ 0.30	& $>0.1$ & 01 & ... & \citet{MAGIC}\\
\hline
3C\;279 &FSRQ &0.536   &   4.20 $\pm$ 0.30	& $>0.1$ & ... & ... & \citet{2019AA...627A.159H}\\	
PKS\;1510-089  &FSRQ &0.361   &   3.97 $\pm$ 0.23	& $>0.15$ & ... & ... & \citet{2018AA...619A.159M} \\	
4C\,+21.35     &FSRQ &0.432   &   3.75 $\pm$ 0.27	& $0.07-0.4$ & ... & ... & \citet{2011ApJ...730L...8A} \\	
S3\,0218+35    &FSRQ &0.954   &   3.80 $\pm$ 0.61	& $0.065-0.175$ & ... & ... & \citet{2016AA...595A..98A} \\	
PKS\;1441+25   &FSRQ &0.939   &   5.3 $\pm$ 0.5	    & $0.08-0.2$ & ... & ... & \citet{2015ApJ...815L..22A}\\	
PKS\;0736+017  &FSRQ &0.189   &   3.1 $\pm$ 0.30 	& $>0.1$ & ... & ... & \citet{2020AA...633A.162H} \\	
B2\;1420+32    &FSRQ &0.682   &   4.22 $\pm$ 0.24	& $>0.1$ & ... & ... & \citet{2020arXiv201211380A}\\
\hline
\end{tabular}
\caption{List of BL Lac objects detected in VHE used in this work.Top group lists the HBL's, Middle group lists the extreme HBL's and bottom group lists the FSRQs. Column description, 1: Source Name 2: Source classification 3: Redshift (z) 4: Observed VHE Index ($\Gamma$) during its lowest flux state 5: VHE spectral range (TeV) 6: Total number of VHE observations available in literature ({\url{http://tevcat.uchicago.edu/}}) 7: Average observed VHE index considering all available observations (with propagated errors) 
7: References of the Low VHE flux state.}

\label{tab1}
\end{table*}

\subsection{Linear Regression Analysis}
\label{sec2.2}
The dependence of the observed VHE spectral index with redshift can be further scrutinized through a linear fit. Such treatments are advantageous in establishing the relation between these two quantities. 
For HBLs, we found the best fit straight line between $\Gamma$ and $z$ to be 
\begin{align}
	\label{eq1}
	\Gamma = (5.34\pm 1.1)z+(2.50\pm 0.12)
\end{align}
with $\chi^2/{\rm d.o.f} = 12.97/22$ and in case of $\Gamma_{av}$ and $z$ 
\begin{align}
    \label{eq2}
	\Gamma_{av} = (5.97\pm 1.15)z+(2.55\pm 0.13)
\end{align}
with $\chi^2/{\rm d.o.f} = 10.99/22$.
In Figure~\ref{Fig1}, we show the best fit line for the case of $\Gamma$ and $z$ as solid green line with 
the residuals in the bottom panel. The residuals are evenly spread over the source redshifts and this advises that a straight
line is capable of explaining the observed trend. We also produced a normal quantile plot to identify the distribution 
of residuals. This also favours a normal distribution of residuals and hence the regression relations given by equation~\ref{eq1} \& \ref{eq2} can be used for prediction.

The analysis is repeated for the case of EHBLs and FSRQs. For EHBLs, the linear regression analysis results are  
\begin{align}
    \label{eq3}
	\Gamma = (5.92\pm 0.9)z+(1.90\pm 0.12)
\end{align}
with $\chi^2/{\rm d.o.f} = 2.52/10$ and 
\begin{align}
    \label{eq4}
	\Gamma_{av} = (5.51\pm 0.88)z+(1.95\pm 0.12)
\end{align}
with $\chi^2/{\rm d.o.f} = 4.41/10$.
In Figure~\ref{Fig2} we show the best fit line with the residuals in the bottom panel for $\Gamma$ and $z$. The low
value of $\chi^2$/d.o.f results from the large errors in the data; however, a straight line is the simplest function that 
can be fitted and the fit statistics can be improved with a larger sample and precise index measurements.
For FSRQs, the linear regression analysis results are
\begin{align}
    \label{eq5}
	\Gamma = (1.82\pm 0.62)z+(3.06\pm 0.33)
\end{align}
with $\chi^2/{\rm d.o.f} = 6.57/5$
and the best fit line is shown in Figure~\ref{Fig3}.

\section{Redshift Estimation}
\label{sec3}
The BL Lac class of blazars often lack emission/absorption line features and this makes it almost impossible to identify the source redshift through optical spectroscopy. Alternatively, the steepening of the VHE spectrum due to the EBL can be used to estimate the source redshift/distance. Usual methods involve prediction of intrinsic VHE spectrum at source through theoretical modelling and assumption of certain EBL model \citep[see e.g.,][]{Mankuzhiyil_2010}. Instead, the regression relations given in section \S \ref{sec2} can also be used to estimate the redshift of blazars. The advantage is the estimated redshifts do not depend either on blazar SED model or the EBL models rather it relies only on the observed correlation between $\Gamma$ and $z$. Below we use the regression relations (equation~\ref{eq1}, \ref{eq2}, \ref{eq3} and \ref{eq4}) to estimate the redshift of five HBLs and one EHBL for which this information is uncertain. The estimated redshifts along with earlier estimates are given in Table~\ref{tab2}.

\subsection*{1ES\,2322-409}
The BL Lac object 1ES\,2322-409 was first detected in VHE by \emph{HESS} during 2004 \citep{10.1093/mnras/sty2686}. A tentative redshift of $0.17359 \pm 0.00018$ for 1ES\,2322-409 was first suggested from the combined redshift and peculiar velocity survey over the southern sky (|b| > $10^{\circ}$), namely 6DF Galaxy survey \citep{10.1111/j.1365-2966.2009.15338.x}. This estimation is based upon low signal-to-noise ratio spectrum and the evidence for the absorption line is weak \citep{10.1093/mnras/sty2686}. On the other hand, if we consider the redshifts of the galaxies obtained from shallow surveys \citep{10.1111/j.1365-2966.2009.15338.x, refId4, 1996ApJ...470..172S, 10.1093/mnras/281.3.L47} which are located close to 1ES\,2322-409, one would estimate the redshift as 0.06. However, it lacks the justification that this source can be a part of the selected group of galaxies.

The VHE observations of 1ES\,2322-409 was reported by \citep{10.1093/mnras/sty2686} and the spectrum can be well described by a 
power law with an index $\Gamma = 3.40 \pm 0.66$ (stat) $\pm$ 0.20 (sys). The photon index falls well within the range of indices used for the correlation study of HBLs (Figure~\ref{Fig1}) and hence the regression relations (equation~\ref{eq1} and \ref{eq2}) can be used to estimate the redshift of 1ES\,2322-409. Using these relations we are able to constrain the source redshift as 0.17 $\pm$ 0.13 and 0.14 $\pm$ 0.12 respectively. Unfortunately, the large error bar in the spectral index reflect as significant uncertainty on the estimated redshift\footnote{Given $m$ and $c$ are the slope and y-intercept of the best fit regression line, the error on redshift, $z=(\Gamma-c)/m$, is estimated as
	\begin{equation*}
		\Delta z = z \sqrt{\left(\frac{\Delta\Gamma}{\Gamma-c}\right)^2 + \left(\frac{\Delta c}{\Gamma-c}\right)^2 + \left(\frac{\Delta m}{m}\right)^2}.
	\end{equation*}}.

\subsection*{H\,1722+119}
H\,1722+119 was identified as a BL Lac object independently by \citet{1989MNRAS.240...33G} and \citet{1990ApJ...350..578B}, and 
its optical spectrum is largely featureless. 
An absorption feature was initially reported by \citet{2011ApJ...729..115A} and estimated the source redshift as 0.018; nevertheless, this was not confirmed by other optical observations \citep{1993A&AS..100..521V, 1993AJ....106...11F, 1994ApJS...93..125F}. Interestingly, observations of the source using ESO's Very Large Telescope did not show any features in its optical spectrum \citep{2006AJ....132....1S} and the authors derived a lower limit of redshift as $z>0.17$. Later using the spectrograph X-shooter of ESO's Very Large Telescope the lower limit of redshift was modified as $z>0.35$ \citep{refId5}. Using \emph{Nordic} Optical Telescope and following the technique described by \citet{Sbarufatti_2005}, \citet{10.1093/mnras/sts410} obtained the redshift lower limit as $z>0.4$. A redshift estimate of z = 0.34 $\pm$ 0.15 using VHE observations by \emph{MAGIC} was put forth by \citep{10.1093/mnras/stw689}. The authors used the fact that the photon index at VHE cannot be harder than the one measured at high energy by \emph{Fermi}. The redshift estimation was done by considering a EBL model by \citet{refId8}.

For the estimation of redshift for H\,1722+119, we use the spectral information obtained 
from the \emph{MAGIC} observations during 17-22 May 2013 for six consecutive nights \citep{10.1093/mnras/stw689}. The VHE
spectrum can be well fitted by a power-law with photon index $\Gamma$ = 3.3 $\pm$ 0.3 (stat) $\pm$ 0.2 (sys). This index corresponds to a redshift of 0.15 $\pm$ 0.07 using equation~\ref{eq1} and 0.13 $\pm$ 0.06 using equation~\ref{eq2}. These values are much less than the redshift lower limits obtained from the featureless optical spectra.
Though the reasons for this deviation are not very clear, probably the intrinsic spectrum is dominated by the non-thermal emission from the jet and hence the source may lack optical line features. The earlier estimate based on VHE study involves extrapolation of the contemporaneous low energy gamma-ray spectrum to VHE and this may be the reason for the difference.

\subsection*{PKS\,1440-389}
PKS\,1440-389 is one of the bright blazars in the \emph{Fermi} energy range and has been classified as high energy synchrotron peaked blazar by the 3rd \emph{Fermi} AGN Catalog \citep{Ackermann_2015}. Based on the 6dF Galaxy Survey \citep{Jones_2004} a redshift of 0.065 was suggested for the blazar PKS\,1440-389. However, due to poor spectral quality this estimate was not included in the 6dF catalog \citep{Jones_2009}. 
Under photo-hadronic interpretation of VHE emission and using EBL model by \citet{refId8}, \citet{2019ApJ...884L..17S} constrained the redshift of PKS\,1440-389 in the range 0.14 $\leq$ z $\leq$ 0.24. Extrapolating the high energy spectrum observed by \emph{Fermi} to VHE energies and using the EBL model by \citet{Dom_nguez_2010}, \citet{10.1093/mnras/staa999} estimated the upper limit of redshift as z $<$ 0.53. In spite of all these measurements at different wavelength, the redshift of PKS\,1440-389 is still uncertain with the best limit being 0.14 $<$ z $<$ 2.2 \citep{Shaw_2013}. 

We considered the VHE spectra index of PKS\,1440-389 from two spectral studies by \emph{HESS} during Feb 2012 \citep{prokoph2015hess, 2020MNRAS.494.5590A}. The mean spectral index from these two observations is 3.66 $\pm$ 0.39. Using the regression relations (equation~\ref{eq1} and \ref{eq2}) this can be translated into the source redshift as 0.21 $\pm$ 0.09 and 0.19 $\pm$ 0.08. These redshift estimates are comparable with earlier values.

\subsection*{PKS\,1424+240}
PKS\,1424+240 is a BL Lac object with a hard high energy spectrum \citep{2009ApJ...707.1310A}. The source was first detected at VHE by \emph{VERITAS} \citep{2010ApJ...708L.100A} and later by \emph{MAGIC} \citep{refIda}. Photometric redshift obtained through spectral fitting of optical/UV data provides an upper limit as $<$ 1.11 \citep{2010ApJ...708L.100A}. A lower limit $>$ 0.6 can be inferred from the Ly$_\alpha$ and Ly$_\beta$ absorption features in far UV spectra \citep{2013ApJ...768L..31F}. Extrapolating the high energy spectra to VHE energies and considering different EBL models, 
\citet{2010ApJ...708L.100A} suggested a redshift upper limit as $<$ 0.66. \citet{2011arXiv1101.4098P} adapted a statistical approach to estimate the redshift of PKS\,1424+240 using the high energy and VHE spectral index for an ensemble of blazars with known redshifts. Their study suggested the redshift of PKS\,1424+240 as 0.24.

For the redshift estimation, we used the two \emph{VERITAS} studies of the source using the VHE observations taken from 2009 to 2011 \citep{2010ApJ...708L.100A, 2015ICRC...34..821B}. The photon spectra can be explained well by a power law and the mean photon index of 4.0 $\pm$ 0.58 has been calculated. Using the mean index and the regression relations (equation~\ref{eq1} and \ref{eq2}), we estimated the source redshift as 0.28 $\pm$ 0.13 and 0.24 $\pm$ 0.11. These are well within the limiting values of redshift and also closely agrees with the estimate of \citet{2011arXiv1101.4098P}.

\subsection*{PG\,1553+113}
PG\,1553+113 is one of the brightest sources known from X-ray to VHE regime located in the Serpens Caput constellation \citep{2002A&A...384...56C}. The optical spectrum of the source is featureless and this makes the redshift estimation difficult through spectroscopic studies. Constraints derived from the apparent magnitude of host galaxy sets a lower limit on redshift as $\geq$ 0.25 \citep{2007AA...473L..17T}. More stringent estimate on redshift, 0.43 $<$ z $<$ 0.58, is obtained through the study of 
far-UV absorption line features arising from the interstellar and intergalactic medium \citep{Danforth_2010}.
Using \emph{HESS} and \emph{MAGIC} observations and assuming the intrinsic photon index cannot be harder than 1.5 along with the EBL model proposed by \citet{2004A&A...413..807K}, \citet{2007ApJ...655L..13M} derive an upper limit on the redshift as $<$ 0.69. \citet{refIds} estimated an upper limit on redshift as $<$ 0.64 from the measured spectral break between the GeV-TeV spectral indices. Assuming the intrinsic VHE spectrum cannot be harder than the high energy and using a Bayesian approach, \citet{Abramowski_2015} found the most probable redshift as $0.49 \pm 0.04$.

We used the VHE spectral information of the source from five distinct observations by \emph{HESS}, \emph{MAGIC} and \emph{VERITAS} 
experiments \citep{2015ApJ...802...65A, 2015ApJ...799....7A, 2011ICRC....8...47B, 2008A&A...477..481A, 2009A&A...493..467A}. 
In all the cases the spectrum is well fitted by a power law and the mean index calculated from these observations is 4.51 $\pm$ 0.26. This corresponds to an redshift of 0.38 $\pm$ 0.10 and 0.33 $\pm$ 0.08 estimated from the regression relations (equation~\ref{eq1} and \ref{eq2}) respectively.

\subsection*{HESS\,J1943+213}

The VHE source HESS\,J1943+213 is identified as EHBL from \emph{HESS} observations \citep{2011}. A lower limit on the redshift z $>$ 0.03 is estimated from the optical spectroscopic studies \citep{refId0p}; however, this is not a stringent condition since the host galaxy was not well defined. Alternatively, from the expected host galaxy flux of a typical BL Lac object, a lower limit on redshift can be obtained as z$>$ 0.14 \citep{2011ICRC....8..109C}. Extrapolating the high energy spectrum observed by \emph{Fermi} to VHE energies and using the EBL model by \citet{refId8}, an upper limit on redshift is obtained as z $<$ 0.45 \citep{refId0p}. Similarly, using \emph{Fermi} and \emph{VERITAS} observations along with the EBL model by \citet{refId8}, \citet{Archer_2018} derived a conservative limit on redshift as $<$ 0.23. A photo-hadronic interpretation of VHE emission placed the limits on redshift as 0.14 $\leq$ z $\leq$ 0.19 \citep{2019ApJ...884L..17S}. The VHE spectral information of HESS\,J1943+213 is obtained from \emph{HESS} observations during 2009 \citep{2011}. The spectrum in the energy range 470 GeV to 6 TeV is well described by a power law with photon index of 3.1 $\pm$ 0.3. Since the source is classified as EHBL, we use the regression relations (equation~\ref{eq3} and \ref{eq4}) for EHBLs and the constraint on redshift can be obtained as 0.20 $\pm$ 0.06 and 0.21 $\pm$ 0.06.

\begin{table*}
\centering
\begin{tabular}{|c|c|c|c|c|c|}
\hline

\multirow{2}{*}{Source Name} & \multirow{2}{*}{Source Type} & \multicolumn{2}{c|}{Redshift Estimate (This Work)} & \multicolumn{2}{c|}{Previous Estimates}\\ \cline{3-6}
& & $\Gamma$ & $\Gamma_{av}$ & Optical based study & Model based VHE study \\
\hline

1ES\,2322-409    &   HBL      &      0.17 $\pm$ 0.13 & 0.14 $\pm$ 0.12 & \makecell{0.17359 $\pm$ 0.00018 \citep{10.1111/j.1365-2966.2009.15338.x} \\ 0.06 (Shallow surveys)} & ... \\
\hline
H\,1722+119      &   HBL     &       0.15 $\pm$ 0.07     & 0.13 $\pm$ 0.06 &   \makecell{0.018 \citep{1989MNRAS.240...33G} \\ $>$0.17 \citep{2006AJ....132....1S} \\ $>$0.35 \citep{refId5} \\ $>$0.4 \citep{10.1093/mnras/sts410}} & 0.34 $\pm$ 0.15 \citep{10.1093/mnras/stw689}\\
\hline
PKS\,1440-389    &   HBL       &      0.21 $\pm$ 0.09     & 0.19 $\pm$ 0.08 &   \makecell{0.065 \citep{Jones_2004} \\ 0.14 $\leq$ z $\leq$ 0.24 \citep{2019ApJ...884L..17S} \\ 0.14 $<$ z $<$ 2.2 \citep{Shaw_2013}} & $<$ 0.53 \citep{10.1093/mnras/staa999}\\

\hline
PKS\,1424+240    &   HBL     &       0.28 $\pm$ 0.13      & 0.24 $\pm$ 0.11 & \makecell{$<$ 1.11 \citep{refId0R} \\ $>$ 0.6 \citep{2013ApJ...768L..31F}} & \makecell{$\approx$ 0.24 \citep{2011arXiv1101.4098P} \\ $<$ 0.66 \citep{2010ApJ...708L.100A}}\\ 
\hline
PG\,1553+113     &   HBL     &       0.38 $\pm$ 0.10  &  0.33 $\pm$ 0.08 & \makecell{$\geq$ 0.25 \citep{2007AA...473L..17T} \\ 0.43 $<$ z $<$ 0.58 \citep{Danforth_2010} \\ 0.49 $\pm$ 0.04 \citep{Abramowski_2015}} & \makecell{$<$ 0.69 \citep{2007ApJ...655L..13M} \\ $<$ 0.64 \citep{refIds}} \\

\hline

HESS\,J1943+213  &   EHBL     &      0.20 $\pm$ 0.06     & 0.21 $\pm$ 0.07 & \makecell{$>$0.03 \citep{refId0p} \\ 0.14 $\leq$ z $\leq$ 0.19 \citep{2019ApJ...884L..17S} \\ $>$ 0.14 \citep{2011ICRC....8..109C}} & \makecell{ $<$ 0.45 \citep{refId0p} \\ $<$ 0.23 \citep{Archer_2018}} \\
\hline
\end{tabular}
\caption{Redshift identification of 6 BL Lac objects along with the previous estimates.
Column description, 1: Source Name 2: Source classification 3: Redshift values estimated in this work using the regression relation 
between (i) $z$ and $\Gamma$, and (ii) $z$ and $\Gamma_{av}$ 
4: Previous redshift estimations based on the (i) optical study and (ii) VHE study along with references. 
}
\label{tab2}
\end{table*}

\section{EBL Model Comparison}
\label{sec4}

Besides estimation of redshifts, the observed $\Gamma$-$z$ correlation can also be used to compare the EBL models. Since direct measurement of EBL is not possible, it is estimated using cosmological models. The techniques employed can be broadly categorised as {\it forward evolution} models, {\it backward evolution} models and some alternate approaches based on the combination of stellar population with cosmic star formation history (section \S \ref{sec1}). To compare whether the cosmological EBL models can explain the observed $\Gamma$-$z$ correlation, we select four EBL models by \citet{Inoue_2013}, \citet{10.1111/j.1365-2966.2012.20841.x}, \citet{Dom_nguez_2010} and \citet{refId8}, which are widely used for VHE spectral study of blazars. Among these models, \citet{Inoue_2013} and \citet{10.1111/j.1365-2966.2012.20841.x} are forward evolution models while, \citet{refId8} is a backward evolution model. The EBL model by \citet{Dom_nguez_2010} differ from these two approaches where it is estimated using the observed evolution of galaxy population over different range of redshifts. 

\begin{figure}
		\centering
		\includegraphics[scale=0.3, angle=270]{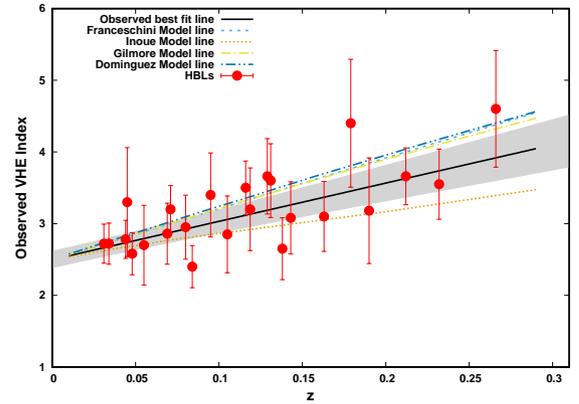}
		\caption{Comparison of observed VHE spectral indices with those predicted by different EBL models for HBLs. Grey region forms the 1-$\sigma$ band on best fit line.}
		\label{Fig5}
\end{figure}

\begin{figure}
		\centering
		\includegraphics[scale=0.3, angle=270]{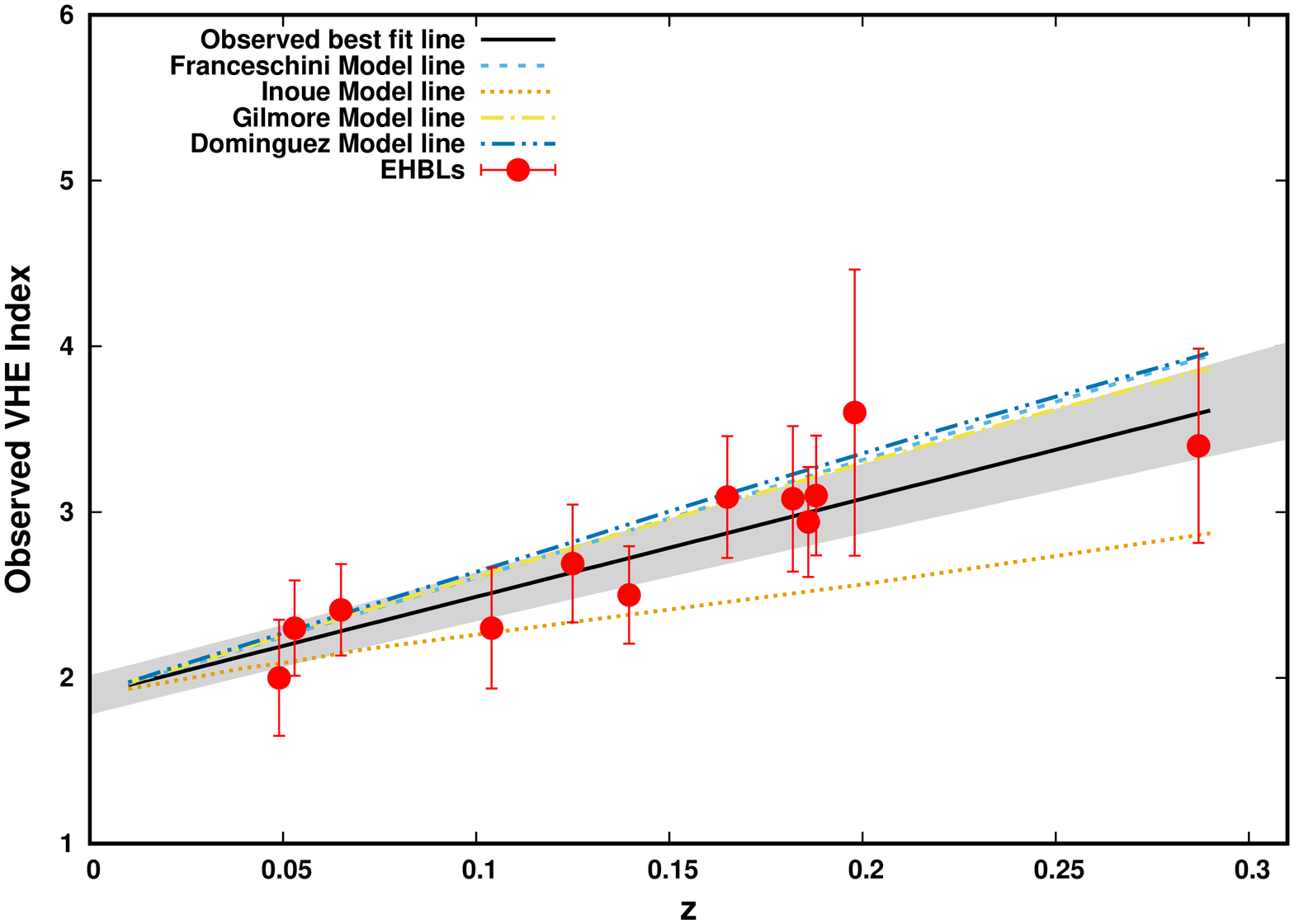}
		\caption{Comparison of observed VHE spectral indices with those predicted by different EBL models for EHBLs. Grey region forms the 1-$\sigma$ band on best fit line.}
		\label{Fig6}
\end{figure}

\begin{figure}
		\centering
		\includegraphics[scale=0.3, angle=270]{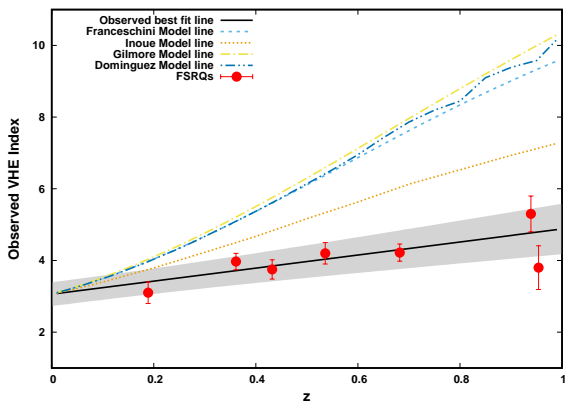}
		\caption{Comparison of observed VHE spectral indices with those predicted by different EBL models for FSRQs.  Grey region forms the 1-$\sigma$ band on best fit line.}
		\label{Fig7}
\end{figure}

The comparison between these models and the observations is performed by predicting the $\Gamma$-$z$ dependence due to these models.
The observed VHE flux $F_o$ at energy $E$ from a blazar is related to its intrinsic source flux $F_i$ as 
\begin{equation}
\label{eq6}
F_o(E) = F_i(E,z) e^{-\tau(E,z)}
\end{equation}
where, $\tau$ is the optical depth due to EBL absorption of VHE photons. If we assume the intrinsic source spectrum to be a power law with index $\Gamma_i$, then the observed spectral slope $\Gamma_o$ at energy $E_*$ will be
\begin{align}
\label{eq7}
\Gamma_o (E_*,z) = \Gamma_i + \left.\frac{d\tau(E, z)}{dln\,E}\right|_{E_*}
\end{align}
The $\tau(E, z)$ for a given $z$ and $E_*$ is estimated using two dimensional linear interpolation of the tabular EBL models and the differentiation is performed numerically. 
For HBLs and EHBLs, $E_*$ is chosen as 1 TeV while it is 300 GeV for FSRQs. This energy is approximately the mean of the observed energies for the sample chosen in this work. 
The intrinsic VHE spectral index $\Gamma_i$ is obtained by extrapolating the regression relations (equation~\ref{eq1}, \ref{eq3} 
and \ref{eq5}) to $z=0$
and the evolution of $\Gamma_o$ is studied against $z$. In Figures~\ref{Fig5} and \ref{Fig6}, we show the predicted $\Gamma_o$ corresponding to the different EBL models for the case of HBLs and EHBLs. The prediction closely satisfy the regression line at small redshifts; however, it starts deviating considerably when the redshift approaches 0.3. Since FSRQs are
detected at higher redshifts compared to HBLs and EHBLs, they are the better sample to perform this study at large distances. 
In Figure~\ref{Fig7}, we show the predicted $\Gamma_o$ corresponding to the different EBL models, for the case of FSRQs, which diverge from the regression line significantly\footnote{It should be noted that the VHE energy at which FSRQs are detected is relatively less compared to HBLs/EHBLs. Accordingly the target EBL photon energy probed will also be different and a direct comparison between $\Gamma_o$ for FSRQs and HBLs/EHBLs is not possible.}.
Even though the number of FSRQs used for this study is less, this result highlights the discrepancy of the EBL models 
in predicting the intrinsic spectrum of high redshift sources. 
The gray band around the regression lines in Figures~\ref{Fig5}, \ref{Fig6} and \ref{Fig7} denote the 1-$\sigma$ error band.

\section{Discussion and Summary}
\label{sec5}
The observed positive correlation between the observed VHE spectral index of blazars with redshift is interpreted as a result of EBL induced absorption effects. Absence of such correlation in low energy bands further strengthen this inference. This led us to develop a novel technique to estimate the redshift of distant blazars whose optical spectrum is featureless. This method neither depend upon the choice of the EBL spectrum nor the intrinsic source VHE spectrum and hence can be treated as model independent/unbiased. 
However, blazar spectral indices are observed to vary significantly during different flux states and this can question the effectiveness of the estimated redshifts. Our study involving 14 \emph{MAGIC} and 8 \emph{VERITAS} observations of Mkn\,421 suggests the standard deviation in the VHE spectral indices to be 0.21 and in case of Mkn\;501 it is 0.14 obtained using 15 \emph{MAGIC} and 3 \emph{VERITAS} observations.

On the contrary, the standard deviation estimated from the whole sample of HBLs considered here is 0.53 which is significantly larger than the variation corresponding to Mkn\,421 and Mkn\,501. In other words, the range of spectral indices introduced by the EBL induced absorption effect is substantial to over shadow the spectral variation of individual sources. Additionally, the spread of 1-$\sigma$ confidence interval of the regression line corresponding to HBLs is 1.1 which is much larger than the spectral variation of individual sources and hence, the associated error on redshift estimation is negligible. 
This is also evident from the minimal variation in the redshift of the source obtained from the regression lines either using $\Gamma$
or $\Gamma_{av}$ (Table~\ref{tab2}). 
Nevertheless, for the source  PG\,1553+113 the variation in its VHE spectral index during different flux states is observed to be very 
minimal \citep{2010ApJ...708.1310A, 2015ApJ...802...65A}.

An important conclusion that can be drawn from the present work is the deviation of the predicted VHE spectral indices due to 
different EBL models from the observed trend. At lower redshifts ($0.06\gtrsim z \lesssim 0.3$), the EBL estimated through forward evolution model by \citet{Inoue_2013} under predicts the spectral indices whereas, the ones estimated from the forward evolution model by \citet{10.1111/j.1365-2966.2012.20841.x}, backward evolution by \citet{refId8} and from the observed evolution of galaxy population \citet{Dom_nguez_2010} over predicts in case of HBL/EHBL class. This probably indicate the missing link between assumed cosmological initial conditions with the present epoch. A correction factor can be introduced as a function of $(z)$ in these EBL models to make them consistent with the regression line. However, this should be scrutinized thoroughly using the acceptable cosmological models and the quantities derived from observations. In case of FSRQs, which can 
probe high redshifts ($z\gtrsim 0.3$), all the EBL models over predicts the VHE spectral indices. However, the target energy at which the spectral index is calculated for the case of FSRQs is different from HBL/EHBL class and hence they probe different energy regimes of EBL (section \S \ref{sec4}). It is interesting to note that the EBL model by \citet{Inoue_2013} under predicts the observed index in case of BL Lacs while it over predicts for the case of FSRQs. Again this can be due to the difference in the EBL energy probed by these studies.
Nevertheless, this study highlights the severe discrepancy between the observations and models. A deeper investigation into this discrepancy involves detailed modelling of cosmological evolution which is beyond the scope of present work.

Most of techniques used to estimate the redshift of blazars through VHE spectral study were based on the prediction of intrinsic 
source VHE spectrum. It is often straight forward to extrapolate the high energy spectrum to VHE energies and consider it as
the upper limit for the intrinsic VHE spectrum. Such studies are capable of providing strong upper limits on the redshifts
of various blazars \citep{Archer_2018, 2010ApJ...708L.100A}. Alternatively, one can consider the emission from the hardest particle distribution that can be obtained through Fermi acceleration process as an approximation of intrinsic VHE spectrum \citep{2006Natur.440.1018A}. This in comparison with the observed VHE spectrum can be used to estimate the redshift under certain EBL model \citep{2007ApJ...655L..13M}. The intrinsic VHE spectrum can also be obtained through broadband SED modelling using synchrotron and SSC emission processes \citep{Mankuzhiyil_2010}. This knowledge can be used to identify the redshift of unknown blazars from their observed VHE spectrum and an appropriate EBL model. This approach of redshift estimation has an additional advantage that the intrinsic VHE spectrum is estimated from the broadband spectral information rather than a narrrow window at high energies. \citet{10.1111/j.1745-3933.2010.00862.x} introduced a different approach to estimate the redshift of blazars by extrapolating of high energy spectrum of blazars with known redshifts ($z$) to VHE energies. Assuming this as the intrinsic VHE spectrum, they estimated the redshift ($z^*$) using the different EBL models. Using a linear regression analysis between $z$ and $z^*$, they were able to predict the redshift of blazars which was not known. However, these techniques have a bias on various assumptions of the intrinsic VHE spectrum and/or the EBL models. The present study, on the contrary, do not have such bias and the redshift estimates can be treated as model independent.

A major assumption in the present approach of redshift estimation is that the intrinsic VHE spectral indices of different classes of blazars are similar. At VHE energies the inverse Compton emission happens mostly at Klein-Nishina regime and the spectral index depends on the emitting electron energies, target photon frequency and the bulk Lorentz factor of the jet \citep{1998ApJ...509..608T}. Hence, this assumption demands that these physical quantities do not vary significantly for a particular class of blazars and the shape/slope of their underlying electron distribution responsible for VHE emission is also similar. The shape/slope of the emitting electron distribution is decided by the acceleration rate and the particle diffusion from the main acceleration region \citep{1998A&A...333..452K, Rieger_2007} and this assumption also implies the particle dynamics in blazar jets are comparable. An alternate explanation for the $\Gamma$-$z$ correlation can be attributed to the cosmological evolution of physical parameters which is particularly effective at VHE energy regime. One such parameter can be the bulk Lorentz factor of the jet which decides the extremity of the Klein-Nishina scattering cross section for the inverse Compton process. However, such studies require rigorous analysis of high quality-simultaneous broadband spectrum of blazars spread over various redshifts.
The present approach of redshift estimation can be largely improved with a tighter correlation between the observed VHE spectral 
index and the redshift. This in turn demands increased number of VHE blazars with precise spectral index measurements. The upcoming 
high sensitivity experiments such as \emph{Cherenkov Telescope Array (CTA)}/\emph{Major Atmospheric Cherenkov Experiment (MACE)} have the potential to achieve this and can be used as an independent tool to estimate the redshift of blazars.

\section{Acknowledgement}
Authors thank the anonymous referee for the valuable comments and suggestions.
M.Z, S.S, N.I \& A.M acknowledge the financial support provided by Department of Atomic energy (DAE), Board of Research in Nuclear Sciences (BRNS), Govt of India via Sanction Ref No.: 58/14/21/2019-BRNS. SZ is supported by the Department of Science and Technology, Govt. of India, under the INSPIRE Faculty grant (DST/INSPIRE/04/2020/002319). This work has made use of the TeV catalog ({\url{http://tevcat.uchicago.edu/}}) created and maintained by Scott Wakely and Deirdre Horan and partially supported by NASA and the NSF.

\section{Data Availability}
The data and the codes used in this work will be shared on the reasonable request to the corresponding author Malik Zahoor (email: malikzahoor313@gmail.com).

\bibliographystyle{mnras}
\bibliography{biblography}

\bsp	
\label{lastpage}
\end{document}